\documentclass[12pt,preprint]{aastex}
\lefthead{Skinner et al.}
\righthead{FU Orionis}

\begin{document}

\title{The Unusual X-ray Spectrum of FU Orionis}

\author{Stephen L. Skinner}
\affil{CASA, Univ. of Colorado, Boulder, CO 80309-0389 }

\author{Kevin R. Briggs and Manuel G\"{u}del}
\affil{Paul Scherrer Institute, W\"{u}renlingen and Villigen,
CH-5235 Switzerland}

%
\newcommand{\ltsimeq}{\raisebox{-0.6ex}{$\,\stackrel{\raisebox{-.2ex}%
{$\textstyle<$}}{\sim}\,$}}
%
\newcommand{\gtsimeq}{\raisebox{-0.6ex}{$\,\stackrel{\raisebox{-.2ex}%
{$\textstyle>$}}{\sim}\,$}}

\begin{abstract}
FU Orionis objects (FUors) have undergone strong optical outbursts
and are thought to be young low-mass stars accreting at high
rates of up to $\dot{\rm M}_{acc}$ $\sim$ 10$^{-4}$ M$_{\odot}$ yr$^{-1}$.
FUors have been extensively studied at
optical and infrared wavelengths, but little is known about their 
X-ray properties. We have thus initiated a program aimed at searching for
and characterizing their X-ray emission. First results are
presented here for the prototype star FU Orionis based on observations
obtained with {\em XMM-Newton}. Its CCD X-ray spectrum
is unusual compared to those of accreting classical T Tauri stars (cTTS).
The cool and hot plasma components typically detected in cTTS are
present but are seen through different absorption column densities.
The absorption of the cool component is consistent with A$_{V}$ $\approx$
2.4 mag anticipated from optical studies but the absorption of the hot
component is at least ten times larger. The origin of the excess absorption
is uncertain  but cold accreting gas or a strong near-neutral wind are
likely candidates.
The hot plasma component accounts for most of the observed X-ray flux and
thermal models give very high temperatures kT $\geq$ 5 keV. The most 
prominent feature in the X-ray spectrum is an  exceptionally strong
Fe K emission line at 6.67 keV and weak emission from fluorescent 
Fe I at $\approx$6.4 keV may also be present. The high plasma 
temperature clearly demonstrates that the emission is dominated
by magnetic processes. We discuss possible origins of the unusual
X-ray spectrum in the context  of a complex physical environment that
likely includes disk accretion,  a strong
wind, magnetic activity, and close binarity.
\end{abstract}




%

\section{Introduction}

FU Orionis objects (FUors) undergo  some of the most
extreme variability seen in low mass pre-main-sequence (PMS) stars. 
The classical FUors are characterized  by optical outbursts during which
the star increases in brightness by several magnitudes
over $\sim$1 - 10 years and then slowly decays   
on timescales of $\sim$20 - 100 years. The underlying
cause of these flare-ups is not known. Herbig (1977)
considered several possible mechanisms but none emerged
as a clear favorite. More recent work
has focused on the idea that the brightenings
are due to a dramatic increase in the accretion
rate through a circumstellar disk, perhaps triggered by
disk instabilities or interactions with a close
binary companion. A review of the FU Orionis phenomenon
and the evidence favoring episodic
accretion as its cause was given by Hartmann \&
Kenyon (1996 = HK96). Potential difficulties with
the accretion interpretation have been discussed
by Petrov \& Herbig (1992).

The sample of classical FUors for which optical outbursts
have actually been observed is small. These include FU Ori
itself, as well as  V1057 Cyg, V1515 Cyg, and  V1735 Cyg.
Because of their unusual variability, the classical FUors
have been extensively studied at optical, infrared, and
submillimeter wavelengths. Their properties are summarized  
by Herbig (1966, 1977), HK96, and Sandell \& Weintraub (2001).
However,  there have been no previous X-ray observations targeted
specifically at classical FUors, and their X-ray properties are largely
unknown.We have thus initiated a program aimed at
searching for and characterizing their X-ray emission. We would
like to know if the X-ray spectra of FUors are similar to
classical T Tauri stars (cTTS). Both FUors and cTTS are accreting
low-mass PMS stars, and the FUor V1057 Cyg was a T Tauri
star before erupting  in 1969 (Herbig 1977).
It is thus believed that FU Ori-like outbursts represent a transient
(and possibly recurrent) phase in the life of a T Tauri star.

We present here first results from X-ray observations of 
the prototype FU Orionis. It brightened optically by 
about 6 magnitudes during 1936 - 1937 (Herbig 1966) and is
still in slow decline. It has an optical spectrum and
colors resembling an F-G supergiant, a near-IR excess,
a high Li abundance, and
strong P Cygni optical and UV line profiles indicative
of mass loss, but no bipolar CO outflow or HH objects 
have been detected (HK96). Features associated with
an expanding circumstellar shell are seen in the 
spectra (Herbig 1966).
A near-IR companion is present at an offset of 
0.50$''$ (230 AU at d = 460 pc) and PA = 161$^{\circ}$
(Wang et al. 2004). This object is about 4 mag fainter 
at K band than FU Ori and has a near-IR excess, so is likely
a PMS star star (Reipurth \& Aspin 2004).
Interferometry observations at H and K probe the star
on AU distance scales, showing evidence for an accretion
disk with $\dot{\rm M}_{acc}$ $\approx$ 10$^{-4}$ 
M$_{\odot}$ yr$^{-1}$ and a mysterious hot spot
located at a projected separation of 10 AU (Malbet et al. 2005). 
The physical origin of the hot spot is uncertain,
but Malbet et al. argue that it may be the signature
of a very close companion. If so, this would make 
FU Ori a triple system.

We find that the X-ray spectrum of FU Ori is quite
unusual compared to what is typically observed in
cTTS. The spectrum does show the cool and hot
components usually seen in cTTS, but the components
are viewed through different absorption columns.
We discuss possible interpretations of the
unusual spectrum and note qualitative similarities with  
the jet-driving T Tauri star DG Tau A.

\section{XMM-Newton Observations}

A short 16 ksec {\em XMM-Newton} observation 
of FU Ori was obtained on 8 March 2004. FU Ori
was clearly detected but the observation
was adversely affected by high background radiation
(Skinner et al. 2005).

We present here the results from a longer
follow-up observation which began on 
2005 April 3 at 15:36 UT and ended at
02:37 UT on April 4.
Data were acquired with the European Photon
Imaging Camera (EPIC), which provides CCD 
imaging spectroscopy from the PN camera
(Str\"{u}der et al. 2001) and two nearly
identical MOS cameras (MOS1 and MOS2;
Turner et al. 2001). The observation was
obtained in full-window mode using the medium
optical blocking filter. The EPIC cameras
provide  energy coverage in the range
E $\approx$ 0.2 - 15 keV with energy 
resolution E/$\Delta$E $\approx$ 20 - 50.
The MOS cameras provide the best on-axis angular 
resolution with FWHM $\approx$ 4.3$''$ at
1.5 keV.

Data were reduced using the {\em XMM-Newton}
Science Analysis System (SAS vers. 6.1).
Event files generated during the standard
processing were filtered to select good
event patterns. Time filters were applied to
remove segments of high background exposure.
This yielded  26.9 ksec of low-background PN exposure
(including 20.9 ksec of contiguous data in the
last half of the observation) and 32.2 ksec
of contiguous exposure per MOS. Spectra and
light curves were extracted from a circular
region of radius  R$_{e}$ = 15$''$ centered
on FU Ori, corresponding to $\approx$68\%
encircled energy at 1.5 keV. Background
spectra and light curves were obtained
from circular source-free regions near
the source (on the same CCD) and were compared with 
background extracted from an annular region around 
the source. The derived results were not 
significantly affected by the region used to
extract background. 
The mean PN background rate in the vicinity of FU Ori
measured during the usable periods of low-background
exposure was 1.09 $\times$ 10$^{-6}$ c s$^{-1}$ arcsec$^{-2}$
(0.5 - 7 keV). 
The SAS tasks
{\em rmfgen} and {\em arfgen} were used to
generate source-specific response matrix
files (RMFs) and auxiliary response files
(ARFs) for spectral analysis. The data were
analyzed using the {\em XANADU} software
package
\footnote{The {\em XANADU} X-ray analysis software package
is developed and maintained by NASA's High Energy
Astrophysics Science Archive Research Center. See
http://heasarc.gsfc.nasa.gov/docs/xanadu/xanadu.html
for further information.},
including {\em XSPEC} vers. 12.2.0.

\section{Results}

\subsection{Spatial and Temporal X-ray Properties}

Figure 1 shows the combined MOS 1 $+$ MOS 2 image of FU Ori 
in the 0.5 - 7 keV band.  
The pipeline processing
detected this source at position (J2000.)
RA = 05$h$ 45$m$ 22.43$s$, Decl. = 
$+$09$^{\circ}$ 04$'$ 12.2$''$, with a formal positional
uncertainty of 0.30$''$.  This position is offset
1.06$''$ from the {\em 2MASS} position of FU Ori
(2MASS J054522.3$+$090412; J = 6.52, H = 5.70, K = 5.16)
and 1.04$''$ from the {\em HST} Guide Star Catalog (GSC) position.
Our own image analysis using data from all three EPIC
detectors gives an X-ray position nearly identical to that
determined by the  pipeline processing but with slightly
smaller offsets of 1.04$''$ from {\em 2MASS} and
0.95$''$ from {\em HST} GSC. These $\approx$1$''$ offsets
are within the positional accuracy expected for 
{\em XMM} EPIC at the signal-to-noise ratio of our data.
We find no other objects within 15$''$ of FU Ori in
the HST GSC 2.2, USNO B1.1 or 2MASS data bases.
Thus, the X-ray emission most likely originates in FU Ori
but a contribution from the faint IR companion located
0.5$''$ to the south is not ruled out since {\em XMM-Newton} 
cannot spatially resolve the two.

Since the X-ray spectrum shows both  cool and hot plasma
components (Sec. 3.2), we have measured the X-ray centroid 
positions in the combined  MOS 1 $+$ MOS 2 image in the soft
0.5 - 2 keV  and harder 2 - 7 keV bands. These measurements
indicate that any offset between the soft and hard band positions
is no larger than $\approx$0.95$''$, which is slightly less than
the 1.1$''$ MOS pixel size. Higher angular resolution observations
will be needed to determine if a sub-arcsecond offset is actually
present.

No  large-amplitude flares are visible in the 
X-ray light curves. The last 20.9 ksec of contiguous
low-background PN exposure gives a  mean background-subtracted 
PN count rate of 7.2 c ksec$^{-1}$ (0.5 - 7 keV; 
R$_{e}$ = 15$''$).
A $\chi^2$ variability test on this 20.9 ksec segment binned
at 2000 s intervals yields
a probability of constant count rate P$_{const}$ = 0.52 
($\chi^2$/dof = 8.1/9). The MOS detectors are less
affected by background flares and thus provide more
usable exposure than the PN, but at lower count rates. 
The summed MOS1 $+$ MOS2 light curve for the last 32.2 ksec
of  exposure has a mean background-subtracted 
rate of  4.9 c ksec$^{-1}$ (0.5 - 7 keV; R$_{e}$ = 15$''$)
and a $\chi^2$ test gives P$_{const}$ = 0.92 ($\chi^2$/dof = 6.7/13)
using 2000 s bins.  

Based on the above analysis, there is no compelling evidence
for X-ray variability in FU Ori during the April 2005 
observation. However,  the EPIC PN fluxes of FU Ori in
this observation were about 50\% larger than in the shorter
March 2004 observation, as determined by the pipeline
processing and our spectral analysis. It thus seems likely
that FU Ori did vary during the $\approx$1 year time interval
between the two observations. Since the inferred variability
is based on a comparison with high-background data acquired
in March 2004, further time monitoring would be useful to 
confirm the variability and constrain its timescale.

\subsection{The X-ray Spectrum of FU Ori}

Figure 2 shows the PN spectrum of FU Ori. 
The spectrum reveals a cool component below $\sim$2 keV
and a hotter component above $\sim$2 keV. Strong line emission
from the Fe K$\alpha$ complex near 6.7 keV is clearly present.
We discuss spectral models and plasma properties
below. \\

\subsubsection{Thermal Emission Models}

Since cool and hot plasma are clearly present, we first
attempted to fit the spectrum with a {\em single-absorption}
two-temperature (2T) optically thin thermal plasma model.
This model assumes that the emission is due to a multi-temperature 
plasma with a cool component  at temperature kT$_{1}$ and hotter
component at kT$_{2}$, where  both components are viewed through the
same absorption column density  N$_{\rm H}$.
We refer to this model in abbreviated notation as 
N$_{\rm H}$$\cdot$(kT$_{1}$ $+$ kT$_{2}$). Spectral fits
with this model can reproduce the hard part of the spectrum
above $\approx$2 keV with a high-temperature plasma 
kT$_{2}$ $\approx$ 7 keV and strong absorption 
N$_{\rm H}$  $\approx$ 8 $\times$ 10$^{22}$ cm$^{-2}$.
However, the overall fit is unacceptable (reduced $\chi^2$ $>$ 1.5)
because the model flux is heavily attenuated below $\approx$1 keV
for reasonable values kT$_{1}$ $\leq$ 1 keV. As a result, the
model fails to account for the emergent flux below  $\approx$1 keV
that is clearly present in the observed spectrum. These results
suggest that the cool plasma below  $\approx$1 keV is viewed
through lower absorption than that inferred above for the hot plasma.

The inability to reproduce the spectrum using a single-absorption
2T model is somewhat surprising because the X-ray spectra of most
TTS can be reasonably well matched with this type of model, 
including TTS in the Orion region (Getman et al. 2005). However, 
it has recently been noted that such a model could not reproduce
the X-ray spectrum of the jet-driving TTS DG Tau A (G\"{u}del et al. 2005).
In that case, an acceptable fit was obtained by allowing
the absorption column density for each plasma component to be
different. We thus attempted to fit the spectrum of FU Ori using
a {\em double-absorption}  model of the form 
N$_{\rm H,1}$$\cdot$kT$_{1}$ $+$ N$_{\rm H,2}$$\cdot$kT$_{2}$. 

The double-absorption model yields very good results.
Table 1 gives best-fit parameters for this model when all abundances
are held fixed at  solar values (Model A) referenced to
Anders \& Grevesse (1989),
and for the case where the Fe abundances of both components are
allowed to vary independently (Model B).
The variable Fe abundance model shown in Figure 2 provides a 
slightly better fit than obtained with solar abundances, but
this is accomplished by increasing the Fe abundance of the hot
component to a value greater than solar in order to reproduce
the strong Fe K line. Figure 3 compares the fits to the Fe K
line obtained with solar abundances and an enhanced Fe abundance.

The temperature and absorption column density derived for the 
cool component from the above double-absorption  models are quite
reasonable. The temperature kT$_{1}$ $\approx$ 0.7 keV ($\approx$8 MK)
is typical of cool plasma detected in most TTS in Orion 
(Preibisch \& Feigelson 2005). Furthermore, the corresponding best-fit 
absorption from models A and B is in the range 
N$_{\rm H,1}$ $\approx$ (4.2 - 5.5) $\times$ 10$^{21}$ cm$^{-2}$,
but with rather large uncertainties (Table 1). The best-fit values
equate to an extinction A$_{\rm V}$  $\approx$ 1.9 - 2.5 mag
using the  N$_{\rm H}$ to A$_{\rm V}$ conversion given by
Gorenstein (1975). 
These values
are in good agreement with that determined from the color excess
E(B $-$ V) = 0.8 for  FU Ori (Herbig 1977) and with the values 
A$_{\rm V}$ = 2.4 mag  (Adams, Lada, \& Shu 1987) and
A$_{\rm V}$ = 1.85 mag (Kenyon, Hartmann, \& Hewett 1988)
derived from fits of its spectral energy distribution.
It thus seems very
likely that we are detecting cool X-ray emission  that
originates in FU Ori and is moderately absorbed by the same 
material that is responsible for the optical extinction.

The temperature inferred for the hot component in the 
above double-absorption  models is  kT$_{2}$ $\approx$ 5.6 - 7.2 keV 
(65 - 83 MK; Table 1). These temperatures are high but  nevertheless 
within the range observed for magnetically-active TTS in Orion 
(Preibisch \& Feigelson 2005). However, the absorption inferred for
this hot component is  much larger than anticipated from optical
extinction estimates. The X-ray spectral fits from Models A and B give
N$_{\rm H,2}$ $\approx$  (8.4 - 12.8) $\times$ 10$^{22}$ cm$^{-2}$,
which corresponds to A$_{\rm V}$  $\approx$ 38 - 58 mag using
Gorenstein (1975).
The nature
and location of the material responsible for the strong X-ray absorption
are uncertain, but the apparent absence of such high absorption 
at visible wavelengths suggests that it is  primarily gaseous.

Although the above double-absorption model provides an acceptable
fit to the spectrum, it may be overly simplistic in one respect.
It assumes that the plasma  viewed through low absorption
is isothermal (kT$_{1}$), and likewise for that viewed through high
absorption (kT$_{2}$). If the low or high absorption components  
(or both) originate in a multi-temperature corona, then it would
be more physically realistic to replace  kT$_{1}$ with a multi-temperature
plasma, and likewise for kT$_{2}$. We return to this issue in
Section 4.3.

\subsubsection{Fe Emission Lines}

The spectrum between 6 - 7 keV is dominated by
strong Fe line emission. The best-fit line
centroid energy in the PN spectrum is 
E$_{line}$ = 6.68 [6.56 - 6.75;
90\% conf.] keV, which is identified with the Fe K
shell complex including Fe XXV. This line emits
maximum power at T$_{max}$ $\approx$ 10$^{7.6}$ K,
providing unambiguous evidence for hot plasma.
The Fe K line flux  accounts for about one-fourth
of the observed flux in the 0.5 - 7 keV band (Table 1). 
A weak feature is also seen near E $\approx$ 7.79 keV 
as a 2$\sigma$ excess in one bin (Fig. 2). The
feature is of low significance but if it is
weak line emission the most likely candidate would
be Ni XXVII at lab energy E$_{lab}$ = 7.806 keV. A broad
emission peak is also present near 0.8 - 1.1
keV. Numerous Fe and Ne lines and the 
O VIII Ly$\beta$ line
occupy this portion of the spectrum
but these lines are not resolved at  EPIC's  spectral 
resolution.

The value of the  Fe abundance for the cool component 
(Fe$_{1}$) determined from the variable abundance model 
is quite uncertain but does converge to values at or
slightly below solar (Model B in Table 1). As mentioned
above, this model requires a high Fe abundance for the 
hot component (Fe$_{2}$) in order to accommodate the 
strong Fe K line. The best-fit Fe$_{2}$  abundance depends
somewhat on the amount of spectral binning used and we obtain 
values in the range Fe$_{2}$ = 2.20 - 2.78 $\times$ solar
(Table 1 Notes). But, 90\% confidence
intervals do allow a value that is barely consistent
with solar. These variable Fe abundance fits constrain
the width of the
Fe K line to its instrumentally-broadened value,
but no significant differences in the Fe$_{2}$
abundance are found if the line width is allowed
to vary. 

If the global metallicity $Z$ is allowed
to vary instead of just Fe, then the derived 
absorption and temperatures are nearly identical
to those of Model B (Table 1 Notes). The metallicity
of the hot component is largely determined by the 
Fe K line and likewise converges to a value above
solar, but is barely consistent with solar at
the lower 90\% confidence bound.

A  weak excess above the continuum
is present at  6.36 [6.30 - 6.42] keV, as can
be seen in  Figure 3. This excess may
be due to weak fluorescent Fe I emission
that would most likely originate in cool neutral 
material being illuminated by the  high-temperature 
X-ray source. If an additional
Gaussian component at 6.36 keV is added to the thermal models
in Table 1, a slight improvement in the fit is 
obtained and the reduced $\chi^2$ values decrease by
5\% - 10\%. Adding the 6.36 keV Gaussian line does not 
significantly affect the best-fit Fe abundance
of the hot component in  the thermal  models. 

In summary, the X-ray spectra do not place a definitive
constraint on the Fe abundance of the cool plasma 
component.  But if the emission is due to an optically
thin plasma then the  Fe abundance of the  hot component
must be near-solar to account for the exceptionally
strong Fe K line.  It seems quite unlikely
that the Fe abundance of the hot component in the FU Ori
spectrum is strongly depleted relative to the solar 
photospheric value, as has been found for some
accreting T Tauri stars (Kastner et al. 2002; Schmitt et al. 2005).
In a more general context, previous studies of active late-type
stars indicate that the coronal Fe abundance is depleted
relative to the solar photospheric value in some stars but
not in others (reviewed by G\"{u}del 2004).

\subsubsection{A Power-law Continuum?}

Because of the unusually hard continuum that is 
clearly present above 2 keV, we have also attempted to
fit the spectrum using a hybrid model consisting of
a cool lightly-absorbed  optically thin thermal plasma plus
a heavily-absorbed power-law  continuum and a Gaussian
Fe K line. The results of this model are given in Table 1
as Model D with components  N$_{\rm H,1}$$\cdot$kT$_{1}$ $+$
N$_{\rm H,2}$$\cdot$(PL $+$ GAUS).

Interestingly, this hybrid model yields a better fit than
that obtained above  with 
two thermal components as measured by a reduced $\chi^2$ that
is $\approx$35\% smaller. As Figure 4 shows, much of this
improvement comes from a tighter fit to the continuum 
above $\approx$2 keV. There is very little
change in the parameters derived for the cool plasma, but
the inferred absorption for the hot plasma N$_{\rm H,2}$ is about a factor
of two below that determined from the purely thermal models.
Even so, N$_{\rm H,2}$ is still an order of magnitude larger 
than expected for FU Ori based on previously published 
A$_{V}$ estimates. 

The width of the Gaussian Fe K line profile was allowed to
vary in this hybrid model. When the PN spectrum is binned to
a mimimum of 10 counts per bin the  best-fit line width 
FWHM = 2.35~$\sigma_{line}$ = 235 eV (Table 1) 
is slightly broader than the value FWHM $\approx$ 160 eV expected for 
an instrumentally-broadened line at 6.7 keV, but the 90\%
confidence range is consistent with no excess broadening.
This fit attempts to account for some of the excess
near 6.4 keV by broadening the Fe K line (Fig. 4 inset) and
thus  overestimates  the true line width. If a second
Gaussian component is added to model the weak excess near
6.4 keV then the best-fit Fe K line width is reduced to
FWHM = 148 eV, consistent with an unbroadened line.
Thus, we find no strong evidence for
non-instrumental broadening at Fe K but higher energy resolution
is needed to obtain a definitive line width measurement.


Thus, from the standpoint of goodness-of-fit, the hybrid
thermal $+$ power-law model (Model D) offers an improvement over 
the purely thermal models (Models A and B). However, either modeling strategy 
provides a statistically acceptable fit. A purely thermal 
model is more straightforward to interpret on physical
grounds  but we do comment further on the possibility of
nonthermal emission in Section 4.5.

\vspace*{0.5in}

\section{Discussion}

\subsection{Origin of the X-ray Emission}

It is well-known that the optical spectrum of FU Ori 
shows complex two-component structure that cannot
be matched by the spectrum of a normal star (Herbig 1966).
We have found that its X-ray spectrum is also complex,
consisting of cool and hot plasma  components viewed
through different absorption columns. We discuss 
possible interpretations below. The cool component 
could be coronal or shock-related, but the  high
temperature T$_{2}$ $\approx$ 65 - 83 MK determined
from thermal fits of the hot component clearly points
to a magnetic origin.

\subsection{The Cool X-ray Component: Corona versus Shocks}

The presence of a cool X-ray component viewed through an
absorption column that is consistent with E(B$-$V) estimates
is not surprising. The inferred temperature kT$_{1}$ $\approx$
0.7 keV (T$_{1}$ $\approx$ 8 MK) is very much in line with
the cool X-ray emission detected in most TTS. The ubiquity
and stable temperature of this component in a large sample of Orion
TTS were noted by Preibisch \& Feigelson (2005). This cool component
is even present in active late-type coronal sources that are
not accreting PMS stars, and thus likely reflects underlying
conditions in the magnetic corona. 

Some ambiguity arises in the above  interpretation
because cool  high-density plasma in an accretion shock could 
masquerade as cool coronal emission in accreting PMS stars such 
as FU Ori. Higher energy resolution X-ray  spectra are needed to 
obtain electron density estimates that can in principle be used 
to distinguish between high-density plasma in an accretion shock
and lower density coronal plasma. However, the interpretation
of  X-ray emission line density diagnostics is not straightforward 
because ultraviolet radiation from the accretion shock 
can alter X-ray line flux ratios, thus mimicking high densities.
In addition, it has recently been argued that soft accretion shock X-rays 
may not be detectable in sources accreting at rates 
$\dot{\rm M}_{acc}$ $\geq$ 10$^{-9}$ yr$^{-1}$ because the shock
is buried too deeply in the star's atmosphere (Drake 2005). The accretion
rate of FU Ori is thought to be several orders of magnitude
above this limit (Malbet et al. 2005), and the ability to detect
accretion-induced X-rays in FU Ori is thus questionable.
Finally we note that even though the temperature of the cool X-ray
component  determined from spectral fits has rather
large uncertainties, the best-fit value kT$_{1}$ $\approx$ 0.7 keV
is about twice as high as  predicted from canonical accretion
shock models using infall speeds of a few hundred 
km s$^{-1}$ (Ulrich 1976). 

Another possibility is that the cool X-ray component originates in
a shocked jet or outflow at some distance from the star. This
interpretation was put forward to explain the soft X-ray
emission of DG Tau A (G\"{u}del et al. 2005), which is known
to drive a jet. However, no jet or strong collimated outflow
has so far been found for FU Ori. But, FU Ori does have a
strong wind and the wind speed inferred from its broad blueshifted 
H$\alpha$ absorption feature is
$v_{\infty}$ $\approx$ 250 - 400 km s$^{-1}$ (Croswell et
al. 1987). If one assumes that a jet is present (but as yet undetected) 
at similar outflow speeds $v$ $\approx$ $v_{\infty}$, then the 
expected X-ray temperature for shocked jet emission is
(Raga et al. 2002; G\"{u}del et al. 2005):~
T$_{s}$ = 1.5 $\times$ 10$^{5}$($v$/100 km s$^{-1}$)$^{2}$~K
$\approx$ 0.94 - 2.4 MK, or kT  $\approx$ 0.08 - 0.21 keV. These shock
temperatures are a factor of $\sim$3 - 8 lower than the best-fit 
values for the cool component kT$_{1}$ given in Table 1,
but are barely consistent with the lower 90\% confidence bound
on kT$_{1}$.

Given the above considerations, we find no strong 
reason at present to favor
an accretion shock or shocked outflow interpretation for the cool 
X-ray emission of FU Ori over the more conventional coronal interpretation.
If higher resolution X-ray spectra and images are 
eventually obtained, the possibility of shock emission  
would be worth reconsidering.

\subsection{On the Possibility of Inhomogeneous Absorption}

It is apparent from the spectral fits discussed above that the
X-ray absorption toward FU Ori is more complex than normally
encountered in T Tauri stars and a physical interpretation
must account for two absorption components. More than one
interpretation is possible because of the limited X-ray spatial
resolution. Since FU Ori is a known multiple system (Sec. 1) it
may be that more than one star is contributing to the observed X-ray
emission. Another possibility  is that we are detecting X-ray
emission from a  multi-temperature plasma  (such as
a corona) viewed through  inhomogeneous absorption. Evidence for
patchy absorption toward FU Ori has previously been mentioned by
Adams, Lada, \& Shu  (1987).

To explore the coronal interpretation, we assume that the 
intrinsic X-ray emission arises in a multi-temperature corona with
an admixture of cool (kT${1}$) and hot (kT${2}$) plasma,
as is commonly the case in magnetically active PMS stars.
Since FU Ori is accreting and is surrounded by a disk 
viewed at an inclination angle $i$ $\approx$ 55$^{\circ}$ 
(Malbet et al. 2005), some of the coronal emission may be
obscured by material close to the star. This could be the
disk or cold gas in the accretion stream. However, at
$i$ $\approx$ 55$^{\circ}$ the entire corona would not
necessarily be obscured. In particular, the polar region
closest to the observer could incur very little disk 
obscuration, assuming a geometrically thin equatorial
disk. And, if a bipolar outflow or jet is present 
(or has been in the past) then a low-density
cavity might be evacuated near the poles (Clarke et al. 2005), 
allowing softer coronal photons to escape.
However, coronal emission from
subpolar latitudes could suffer heavy disk obscuration or
strong absorption from cold accreting gas that impacts 
the star at subpolar latitudes. 

We represent the above picture with a model of the form
N$_{\rm H,1}$$\cdot$(kT$_{1}$ $+$  kT$_{2}$) $+$
N$_{\rm H,2}$$\cdot$(kT$_{1}$ $+$  kT$_{2}$). This model
accounts for the multi-temperature structure that is 
expected for coronal plasma and assumes that a fraction
of the coronal emission is viewed through relatively
low absorption (N$_{\rm H,1}$) and the remainder through
high absorption (N$_{\rm H,2}$).

The above  model is referred to as Model C in Table 1.
It has four normalization parameters. The low absorption
component has a normalization parameter for the cool
plasma norm$_{1,cool}$ and for the hot plasma norm$_{1,hot}$.
Likewise, the high-absorption component has
norms norm$_{2,cool}$ and norm$_{2,hot}$. When fitting the
X-ray spectrum, all norms are allowed to vary independently
except for norm$_{2,cool}$ which is constrained to be
norm$_{2,cool}$ = [norm$_{1,cool}$/norm$_{1,hot}$]$\times$norm$_{2,hot}$.
That is, we require the ratio of the norms, or equivalently, 
the emission measures, of the cool to hot plasma be the same 
in both the low and high absorption components. This
would be expected if the emission seen through low and
high absorption originates in the same structure (e.g.
a multi-temperature corona). We choose to constrain
norm$_{2,cool}$ because this component is the least
well-determined observationally. Almost all softer
X-ray photons from the cool plasma component that
pass through the high-absorption column will be absorbed
and thus absent from the observed spectrum.

As the results in Table 1 show,  Model C gives a
slightly better fit to the PN spectrum than do Models 
A or B. Even so, there is very little change in the
derived absorption, temperatures, and Fe abundance.
The primary difference between Model C and Models A/B 
is in the underlying physical picture. Model C is 
based on the premise that the observed spectrum 
arises from viewing a single multi-temperature source
through an inhomogeneous absorber. Thus, the emission
detected through low absorption is an admixture of
cool and hot plasma, as is that detected through high
absorption. Another notable difference is that the
X-ray luminosity inferred from Model C is higher than
that of the other models. Nevertheless, the value
L$_{\rm x}$ = 1.7 $\times$ 10$^{31}$ ergs s$^{-1}$ (0.5 - 7 keV)
from Model C is within the range observed for other low-mass PMS
stars in Orion (Getman et al. 2005).

The relative contributions of cool and hot plasma to
the total emission measure derived from Model C are 
somewhat uncertain, but the spectral fits suggest that
they could be roughly equal. However, the total  emission 
measure of the plasma viewed through high absorption
is at least an order of magnitude larger than that 
detected through low absorption. Thus, if the emission
does arise in a multi-temperature corona then the 
corona is almost totally obscured and only a small
fraction of the softer coronal emission manages to
escape. It is not yet clear what would create the escape
path but clumpy absorption, scattering, or a
low-density cavity evacuated by an existing or
fossil jet or collimated outflow are possibilities.

The column density of the heavily-absorbed component 
determined from Model C is
N$_{\rm H,2}$ = 10.6 $\times$ 10$^{22}$ cm$^{-2}$.
This includes the small contribution  N$_{\rm H}$
$\approx$ 0.4 $\times$ 10$^{22}$ cm$^{-2}$ that 
is known to be present based on the optical extinction. 
The excess absorption of the hot component is then
N$_{\rm H,2}^{(excess)}$ = 10.2 $\times$ 10$^{22}$ cm$^{-2}$.
If the accretion column is responsible for this excess,
then we can take the minimum inner disk radius
r$_{min}$ = 5.5 R$_{\odot}$ = 3.8 $\times$ 10$^{11}$ cm
(Malbet et al. 2005) as the approximate length of the 
absorption column. Realistically, this value should be
interpreted as an upper limit because viewing geometry
considerations make it unlikely that we
are looking through the entire length of the accretion
column. Thus, if the strong absorption is due to 
accreting gas then the mean  number density in the 
accretion column is 
n$_{\rm H}$ $\geq$ 3 $\times$ 10$^{11}$ cm$^{-3}$.

\subsubsection{Wind Absorption}

In the above, it was assumed that the accretion flow
is primarily responsible for the heavy X-ray absorption. 
However, the strong stellar wind of FU Ori may also
contribute. Radiative transfer models that assume
rapid wind acceleration give a  high mass loss rate
for FU Ori of 
$\dot{\rm M}$ $\sim$ 10$^{-5}$ M$_{\odot}$ yr$^{-1}$
and P-Cygni type H$\alpha$ absorption features imply
a terminal wind speed 
v$_{\infty}$ $\approx$ 250 - 400 km s$^{-1}$
(Croswell et al. 1987).
There is some evidence that the
wind arises from the surface of the accretion disk
(Calvet, Hartmann, \& Kenyon 1993).

A {\em VLA} radio observation did not detect FU Ori
with a 3$\sigma$ upper limit on its 3.6 cm flux density
S$_{3.6}$ $\leq$ 0.05 mJy (Rodriguez, Hartmann, \&
Chavira 1990). 
Assuming a spherically-symmetric wind with  
temperature T$_{wind}$ $\approx$ 5000 K
(Croswell et al. 1987), v$_{\infty}$ = 400 km s$^{-1}$,
and a distance of 460 pc,
the radio non-detection gives an upper limit on the
{\em ionized} mass-loss rate
$\dot{\rm M}_{ion}$ $\leq$ 4.4 $\times$ 10$^{-8}$ M$_{\odot}$ yr$^{-1}$
(eq. [7] of Skinner, Brown, \& Stewart 1993). Thus, the
high mass-loss rate derived by Crosswell et al. (1987)
can only be reconciled with the radio data if the
wind is largely neutral.

The predicted neutral hydrogen column density along the
line-of-sight toward the star for a 
spherically-symmetric homogeneous wind with mass
loss rate $\dot{\rm M}$ = 10$^{-5}$ M$_{\odot}$ yr$^{-1}$,
terminal speed v$_{\infty}$ = 400 km s$^{-1}$, and
stellar radius R$_{*}$ $\approx$ 4 R$_{\odot}$ 
(Calvet et al. 1993) is
N$_{\rm H,wind}$ $\sim$ 10$^{24}$ cm$^{-2}$
This is about ten times larger than that determined
for the heavily-absorbed component in the X-ray 
spectrum. The X-ray data could be brought into 
agreement with the predicted N$_{\rm H,wind}$
if the mass-loss rate were about an order of 
magnitude smaller, that is 
$\dot{\rm M}$ $\sim$ 10$^{-6}$ M$_{\odot}$ yr$^{-1}$.
The mass loss rate determined from radiative 
transfer models is known to be sensitive to
the poorly-known wind temperature and wind
velocity profile, and if the wind is slowly
accelerated then lower mass loss rates are possible 
and $\dot{\rm M}$ $\sim$ 10$^{-6}$ M$_{\odot}$ yr$^{-1}$
is not out of the question (Croswell et al. 1987).

If the strong X-ray absorption is indeed due to 
FU Ori's massive wind, then an additional question
arises.  Where does the cool component in the X-ray
spectrum originate? If it is cool coronal emission
that originates close to the star, then it must 
somehow escape through the wind without being 
totally absorbed. This would seemingly require
either an inhomogeneous wind or a non-spherical
wind geometry that allows some soft coronal photons
to reach the observer. Here, it is worth mentioning
that a slowly accelerating non-spherical collimated flow was 
mentioned by  Croswell et al. as a possible solution to 
matching the observed optical line properties of
FU Ori, including the absence of redshifted emission.
On the other hand, if the simplistic picture of a 
spherically-symmetric homogenous wind is approximately 
correct then a caculation of the radius of X-ray
optical depth unity in the wind shows that the cool
X-ray emission must emerge far from the star.

\subsection{Issues Concerning Binarity}

It may be that  more than
one X-ray source is contributing to the observed spectrum.
This is conceivable since FU Ori is known to be 
a double or perhaps even a triple system. If each component has
a disk and the disks are viewed at different inclination 
angles (or if one star is viewed through the disk of 
another), then multiple absorption components would be expected.

It seems unlikely that the hard heavily-absorbed X-ray
component originates in the near-IR companion 
0.5$''$ away if its extinction is only  
A$_{\rm V}$ $\approx$ 1.1 mag (Wang et al. 2004).
The extinction toward the hard X-ray component is
much higher (Sec. 3.2.1). Assuming that the 
extinction toward the near-IR companion has not been
underestimated, we are left with the intriguing
possibility that the hard X-rays arise in a third
heavily-obscured magnetically-active component
lying very close to FU Ori. This component 
could be related to the hotspot detected in
the interferometer observations of  Malbet et al. (2005),
or perhaps an object embedded in the disk of FU Ori. It has
been suggested that the periodic   H$\alpha$ emission of
FU Ori might be induced by a low-mass protostar or protoplanet
in the disk (Vittone \& Errico 2005).

Binarity might at first glance seem to provide an attractive
explanation for the unusual double-absorption spectrum of
FU Ori. But, the binary hypothesis is not a panacea. Work
in progress indicates that the cTTS GV Tau A also has a 
double-absorption X-ray spectrum, but a {\em Chandra}
observation shows that a  known protostellar 
companion does not contribute to the X-ray emission 
(G\"{u}del et al., in preparation). Unless an additional
companion is present much closer to the star, 
these results suggest that
other factors besides binarity will be needed to fully
explain the double absorption spectra.

\subsection{Comments on Nonthermal X-ray Emission}

The improvement in the fit to the hard component of the
FU Ori spectrum that results from  replacing the optically 
thin plasma model with a power-law $+$ Gaussian Fe K
line model is substantial (Table 1). This result is 
somewhat surprising, but some caution is warranted
since it is based on the analysis of a relatively low 
signal-to-noise CCD spectrum.

Although hard power-law X-ray continua are often seen in
strongly-accreting compact objects, there is little observational
support to date for power-law X-ray emission from magnetically
active late-type stars (G\"{u}del 2004). A possible exception
is AB Dor, for which a nonthermal X-ray 
continuum excess  was postulated to explain a weak Fe K line
in the presence of X-ray flares (Vilhu et al. 1993). 

However, the situation encountered for FU Ori is different
than for AB Dor.  There is no evidence
for strong X-ray flaring in the FU Ori light curve and the 
Fe K emission line is unusually strong, rather than weak.
It is thus difficult to argue that any existing power-law 
emission is related to large X-ray flares.
As mentioned above (Sec. 4.3.1), a previous {\it VLA} radio observation 
failed to detect FU Ori at 3.6 cm down to rather low limits. 
Thus, we have no radio evidence for
the existence of a population of nonthermal particles in the 
magnetosphere. Even so, additional centimeter radio observations at 
longer wavelengths might be worthwhile since nonthermal radio 
flux density  typically increases with wavelength. 

If a power-law component is present, then magnetic accretion might provide
an alternative explanation. Lamzin (1999) has discussed the
possibility of  nonthermal X-ray emission in accreting PMS
stars. If the density of the infalling gas 
drops below $n$ $\approx$ 10$^{11}$ cm$^{-3}$, then protons can gyrate
many times around a magnetic field line before a collision, and
the accretion shock  passes into the collisionless regime.
In that case, the particle distribution is non-Maxwellian and 
excess hard  X-ray emission can be produced.

At present, the absence of strong X-ray flares and the lack of
a radio detection of FU Ori, along with the strong Fe K line,
suggest that the X-ray emission is predominantly thermal. 
A higher signal-to-noise spectrum will be needed to distinguish
between purely thermal models and models that invoke a hard 
power-law continuum.

\section{Summary}

The double-absorption X-ray spectrum of FU Ori is unusual
compared to most TTS, but in some respects does resemble that
of the jet-driving cTTS DG Tau A (G\"{u}del et al. 2005).
However, there are also notable differences. The spectrum
of DG Tau A does not show a strong Fe K line as is present
in FU Ori, and there is no evidence to date that FU Ori 
has a jet-like outflow.  
If the cool low-absorption X-ray emission detected in DG Tau A
is related to a shocked jet (G\"{u}del et al. 2005) then it is 
not obvious that this interpretation applies to FU Ori as well.

In general, the existing data give
little reason to invoke shock-induced X-rays  to
explain the emission in the FU Ori spectrum, 
despite the belief that it is
accreting at high rates. Clearly, higher angular resolution
observations would be useful to determine if the soft 
emission is slightly offset from the star, as might be the
case if it originates in a shocked jet or outflow.
A higher angular resolution {\em Chandra} observation might also shed
light on whether the faint near-IR companion located
0.5$''$ south of FU Ori contributes to the X-ray emission.

The overall temperature structure of  the FU Ori
spectrum strongly resembles that seen in active late-type
coronal sources, including magnetically-active PMS stars. 
Thus, a coronal origin seems likely, at least for the
hot plasma component.  Assuming that the emission is
coronal, then the nature and location of the material
responsible for the strong absorption of the hard
X-ray component remains the most intriguing question. 
Either cold accreting gas or a strong neutral wind are
the most likely candidates, but an inhomogenous 
absorber or non-spherical geometries are needed if
both the cool and hot plasma originate close to the star.

Even though FU Ori is the prototype,
it would be premature to conclude that its unusual X-ray properties
are representative of the class of FUors as a whole. Prototypes
can show extreme or unusual behavior and T Tauri itself is a
well-known example. Additional  observations of other
FUors are needed to define the X-ray properties of the class
and such observations are now pending.


\acknowledgments

This research was supported by NASA  grants 
NNG04GH23G and NNG05GJ15G. Work at PSI (MG)
was supported by the Swiss National Science
Foundation, grant number 20-66875.01. This work 
is based on observations obtained with 
{\em XMM-Newton}, an ESA science mission with
instruments and contributions directly funded
by ESA member states and the USA (NASA).

\clearpage

\begin{deluxetable}{lllll}
\tabletypesize{\scriptsize}
\tablewidth{0pc}
\tablecaption{{\em XMM-Newton} Spectral Fits for FU Ori 
   \label{tbl-1}}
\tablehead{
\colhead{Parameter}      &
\colhead{ }        &
\colhead{  }
}
\startdata
Model\tablenotemark{a}              &           A         &         B           &        C            &       D                  \nl
Emission                            & thermal             & thermal             & thermal             & thermal $+$ power-law $+$ line    \nl
Abundances                          & solar               & Fe varied           & Fe varied           & solar             \nl
N$_{\rm H,1}$ (10$^{22}$ cm$^{-2}$) & 0.55 [0.02 - 1.13]  & 0.42 [0.00 - 1.81]  & 0.35 [0.12 - 0.90]  & 0.35 [0.23 - 1.11] \nl
kT$_{1}$ (keV)                      & 0.65 [0.17 - 0.86]  & 0.67 [0.08 - 1.01]  & 0.68 [0.14 - 1.04]  & 0.69 [0.08 - 1.03] \nl
norm$_{1}$ (10$^{-6}$)\tablenotemark{b} & 6.22 [1.76 - .....] & 4.68 [0.49 - 27.1]  & 4.87 [0.36 - .....]\tablenotemark{d} & 3.30 [0.90 - 8.52]  \nl 
N$_{\rm H,2}$ (10$^{22}$ cm$^{-2}$) & 12.8 [8.25 - 20.2]  & 8.44 [5.58 - 13.7]  & 10.6 [6.76 - 18.6]  & 4.38 [1.62 - 8.76] \nl
kT$_{2}$ (keV)                      & 5.58 [3.99 - 9.01]  & 7.17 [4.90 - 10.2]  & 7.04 [4.86 - 10.3]  & ...                 \nl
norm$_{2}$  (10$^{-5}$)\tablenotemark{b} & 18.4 [12.2 - 27.2]  & 10.6 [6.70 - 16.4]  & 29.8 [12.5 - 50.6]\tablenotemark{e} & 0.62 [0.31 - 1.99] \nl
$\Gamma$                      & ... & \nodata             & ...                                       & 0.72  [0.07 - 1.61] \nl
E$_{line}$ (keV)              & ... & \nodata             & ...                                       & 6.68  [6.56 - 6.75]    \nl
$\sigma_{line}$ (keV)         & ... & ...                 & ...                                       & 0.10  [0.06 - 0.25]\tablenotemark{g}  \nl
norm$_{line}$ (10$^{-6}$)     & ... & ...                 & ...                                       & 2.58  [1.60 - 3.86]  \nl     
Fe$_{1}$                          & \{1.0\}               & 0.87 [0.00 - .....]\tablenotemark{c,f}      & \{1.0\}   & \{1.0\}        \nl
Fe$_{2}$                          & \{1.0\}               & $\geq$1.0\tablenotemark{c,f}  & 2.52 [1.24 - 4.83]   & ...                 \nl
$\chi^2$/dof                    & 27.3/26                 & 21.6/24                   & 21.0/24       & 13.6/24             \nl
$\chi^2_{red}$                  & 1.05                    & 0.90                      & 0.88          & 0.56 \nl
F$_{\rm X}$ (10$^{-14}$ ergs cm$^{-2}$ s$^{-1}$)          & 7.93 (31.3) & 8.57 (21.9) & 8.58 (68.7)   & 8.49 (12.6) \nl
F$_{\rm X,1}$ (10$^{-14}$ ergs cm$^{-2}$ s$^{-1}$)        & 0.37 (1.58) & 0.34 (1.06) & 0.58 (1.11)   & 0.31 (0.82) \nl
F$_{\rm X,line}$ (10$^{-14}$ ergs cm$^{-2}$ s$^{-1}$)     & 1.40 (1.69) & 2.05 (2.32) & 1.95 (2.28)   & 2.33 (2.48) \nl
L$_{\rm X}$ (10$^{30}$  ergs s$^{-1}$)                    & 7.91        & 5.55        & 17.4          & 2.99 \nl
log [L$_{\rm X}$/L$_{bol}$]                               & $-$5.23     & $-$5.38     & $-$4.89       & $-$5.65     \nl
\enddata
\tablecomments{
Based on  XSPEC (vers. 12.2.0) fits of the background-subtracted EPIC PN spectrum binned 
to a minimum of 10 counts per bin using 26.9 ksec of low-background exposure. Thermal
emission was modeled with the $vapec$ optically thin plasma model in XSPEC.  
The tabulated parameters
are absorption column density (N$_{\rm H}$), plasma energy (kT),
component normalization (norm), photon power-law index ($\Gamma$), Gaussian line
centroid energy (E$_{line}$), line width ($\sigma_{line}$ = FWHM/2.35),
and Fe abundance relative to the solar photospheric value.
Solar abundances are referenced to  Anders \& Grevesse (1989).
Square brackets enclose 90\% confidence intervals and an ellipsis means that 
the algorithm used to compute confidence intervals did not converge.
Curly braces \{...\} enclose quantitities that were held fixed during fitting. 
The total X-ray flux (F$_{\rm X}$) and flux of the low-absorption component
(F$_{\rm X,1}$) are the absorbed values in the 0.5 - 7 keV range, followed in 
parentheses by  unabsorbed values. The continuum-subtracted Fe K line flux
(F$_{\rm X,line}$) is measured in the 6.5 - 6.84 keV range.
The unabsorbed luminosity L$_{\rm X}$ (0.5 - 7 keV)  assumes a
distance of 460 pc. A value L$_{bol}$ = 350 L$_{\odot}$ is adopted
based on an average of values given in the literature (HK96, Levreault 1988, 
Sandell \& Weintraub 2001, Smith et al. 1982). }
\tablenotetext{a}{Model A and B:~N$_{\rm H,1}$$\cdot$kT$_{1}$ $+$ N$_{\rm H,2}$$\cdot$kT$_{2}$;
Model C:~N$_{\rm H,1}$$\cdot$(kT$_{1}$ $+$ kT$_{2})$  $+$ N$_{\rm H,2}$$\cdot$(kT$_{1}$ $+$ kT$_{2}$); \\
Model D:~N$_{\rm H,1}$$\cdot$kT$_{1}$ $+$ N$_{\rm H,2}$$\cdot$(PL $+$ GAUSS)}
\tablenotetext{b}{For thermal $vapec$ models, the norm is related to the emission measure (EM) by
                  EM = 4$\pi$10$^{14}$d$_{cm}^2$$\times$norm, where d$_{cm}$ is the stellar
                  distance in cm.}
\tablenotetext{c}{The Fe abundance of the cool plasma component (Fe$_{1}$) is not
                  well-constrained by the data. 
                  The derived iron abundance of the hot plasma component (Fe$_{2}$) is 
                  largely determined by the fit to the Fe K emission line and is moderately
                  sensitive to the amount of spectral binning. When binned to a minimum of
                  10 counts per bin the derived value is Fe$_{2}$ = 2.78 [1.40 - 5.29; 90\% conf.]
                  $\times$ solar.
                  At a minimum of 15 counts per bin the derived value 
                  is Fe$_{2}$ = 2.20 [0.98 - 4.19].}
\tablenotetext{d}{The quoted value norm$_{1}$ is the total normalization factor for the low-absorption component,
                  which is the sum of the norms for cool and hot plasma:
                  norm$_{1}$ = norm$_{1,cool}$ $+$ norm$_{1,hot}$. The best-fit gives 
                  norm$_{1,cool}$/norm$_{1,hot}$ = 1.59}
\tablenotetext{e}{The quoted value norm$_{2}$ is the total  normalization factor for the high-absorption component:~
                  norm$_{2}$ = norm$_{2,cool}$ $+$ norm$_{2,hot}$. The value of norm$_{2,cool}$ is constrained during the 
                  fit to   norm$_{2,cool}$ =  [norm$_{1,cool}$/norm$_{1,hot}$] $\times$ norm$_{2,hot}$.} 
\tablenotetext{f}{If the global metallicity $Z$ is allowed to vary instead of just Fe alone,
                  then the best fit N$_{\rm H}$ and kT values change by less than 7\%.
                  The best-fit metallicities are $Z_{1}$ = 0.99 [0.01 - 1.61] and
                  $Z_{2}$ = 3.56 [1.04 - 5.00].} 
\tablenotetext{g}{If a second Gaussian component is added at E$_{line}$ = 6.36 keV
                  to model the weak excess that may be due to Fe I, then the width of
                  the Fe K line converges to a value $\sigma_{line}$ = 0.063 [0.00 - 0.20] keV
                  that is consistent with no excess broadening beyond instrumental.}
\end{deluxetable}

\clearpage

\begin{figure}
\figurenum{1}
\epsscale{1.0}
\includegraphics*[width=8.5cm,angle=0]{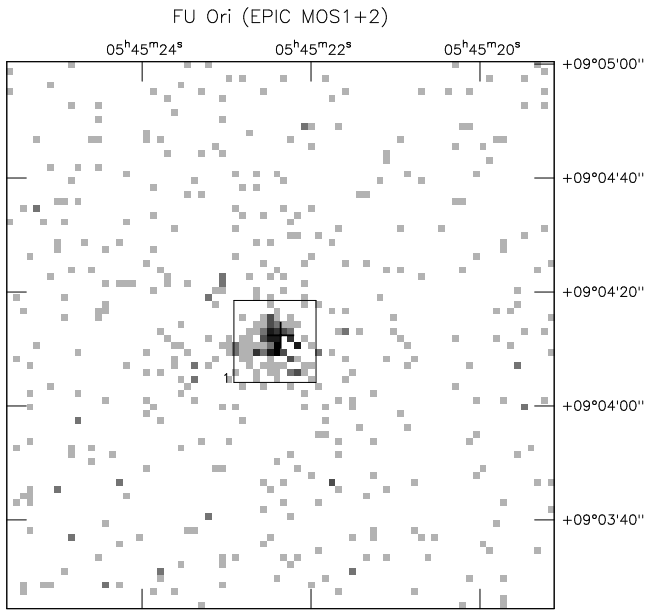}
\caption{
Combined EPIC MOS1 $+$  MOS2 image of FU Ori
(172 net counts, 32.2 ksec per MOS) in the 0.5-7 keV range.  
Cross marks 2MASS position of FU Ori. 
Pixel size is 1.1$''$ and coordinates are  J2000. 
}

\end{figure}

\clearpage

\begin{figure}
\figurenum{2}
\epsscale{1.0}
\plotone{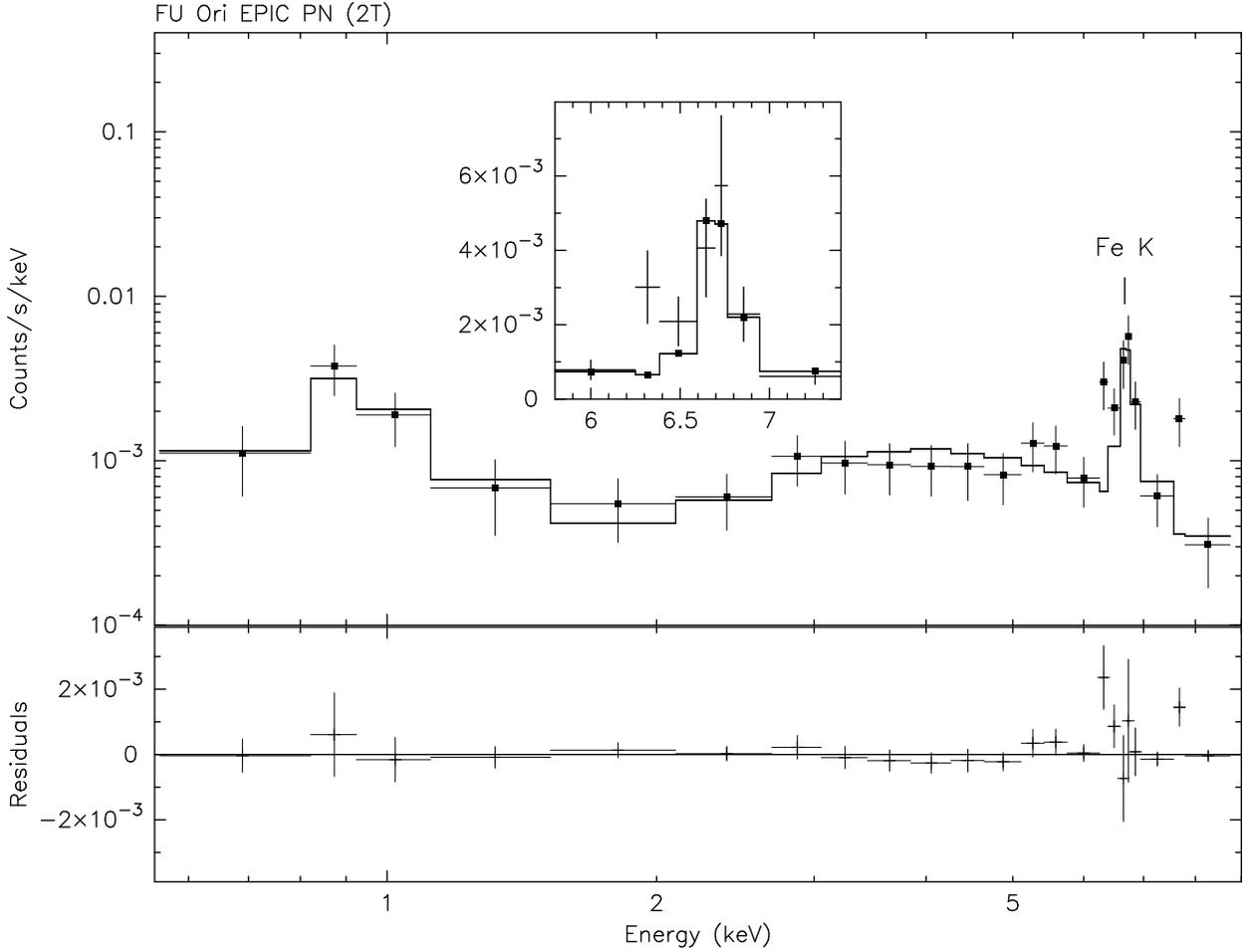}
\caption{
Background subtracted EPIC PN  spectrum of FU Ori
obtained on 2005 Apr 3-4 using  26891 s of low-background
exposure ($\approx$200 net counts in the 0.5 - 7 keV range).
The spectrum (filled squares) is binned to a minimum
of 10 counts per bin. The overlaid model (solid line) is a
double-absorption model with two thermal plasma components
and variable Fe abundance (Model B in Table 1). The Fe K
line fit includes only instrumental broadening. The inset shows
the fit to the Fe K line on a linear  axis scale.
}
\end{figure}

\clearpage

\begin{figure}
\figurenum{3}
\epsscale{1.0}
\plotone{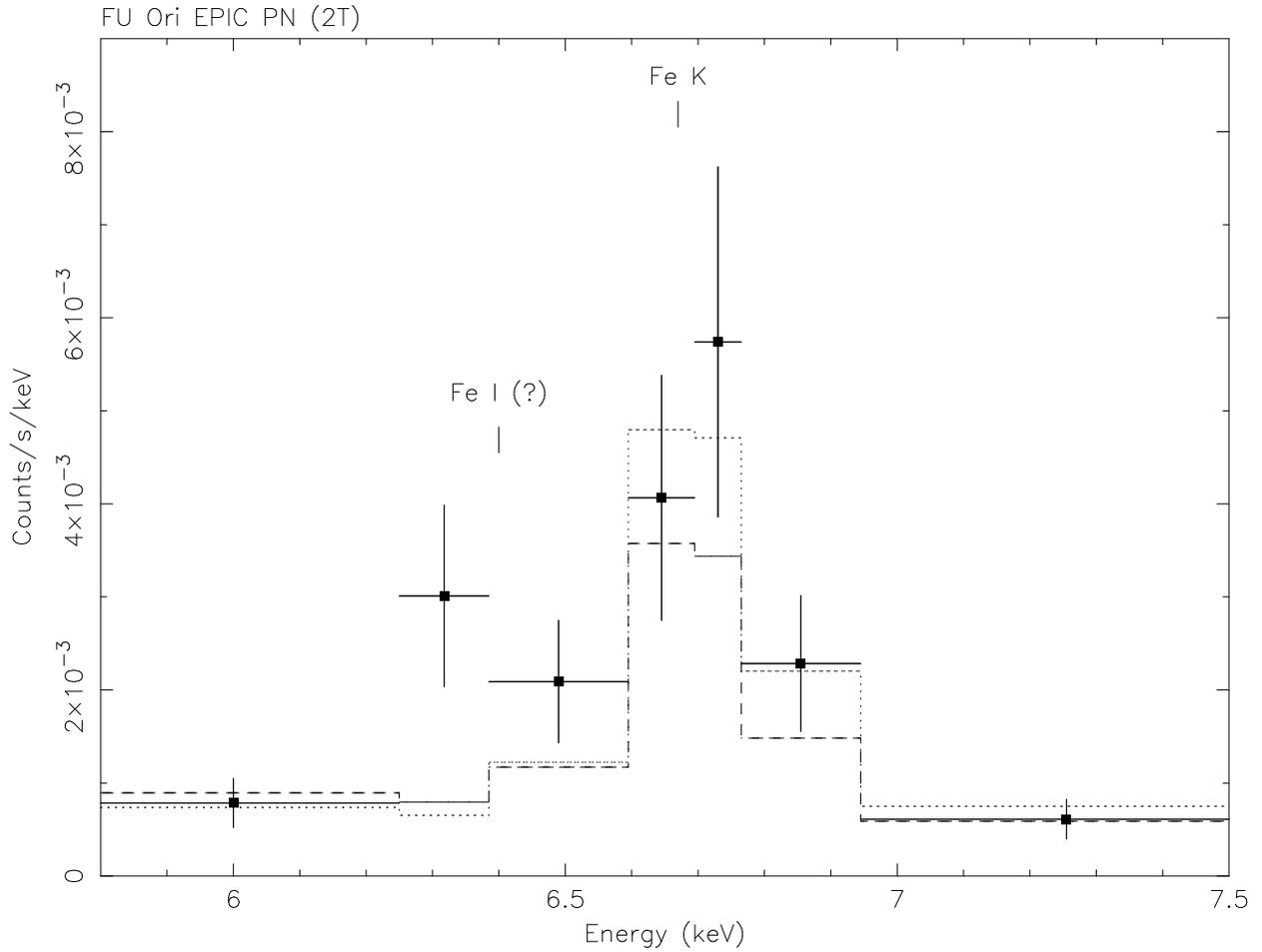}
\caption{
         Fits of the Fe K line in the EPIC PN spectrum of 
         FU Ori with a double-absorption  thermal plasma model.  
         The dashed line shows the best-fit
         using solar abundances (Model A in Table 1) and the 
         dotted line shows the best-fit when the Fe abundances
         are allowed to vary (Model B.).
         The excess emission near 6.4 keV may be due to
         fluorescent Fe I.
}
\end{figure}

\clearpage

\begin{figure}
\figurenum{4}
\epsscale{1.0}
\plotone{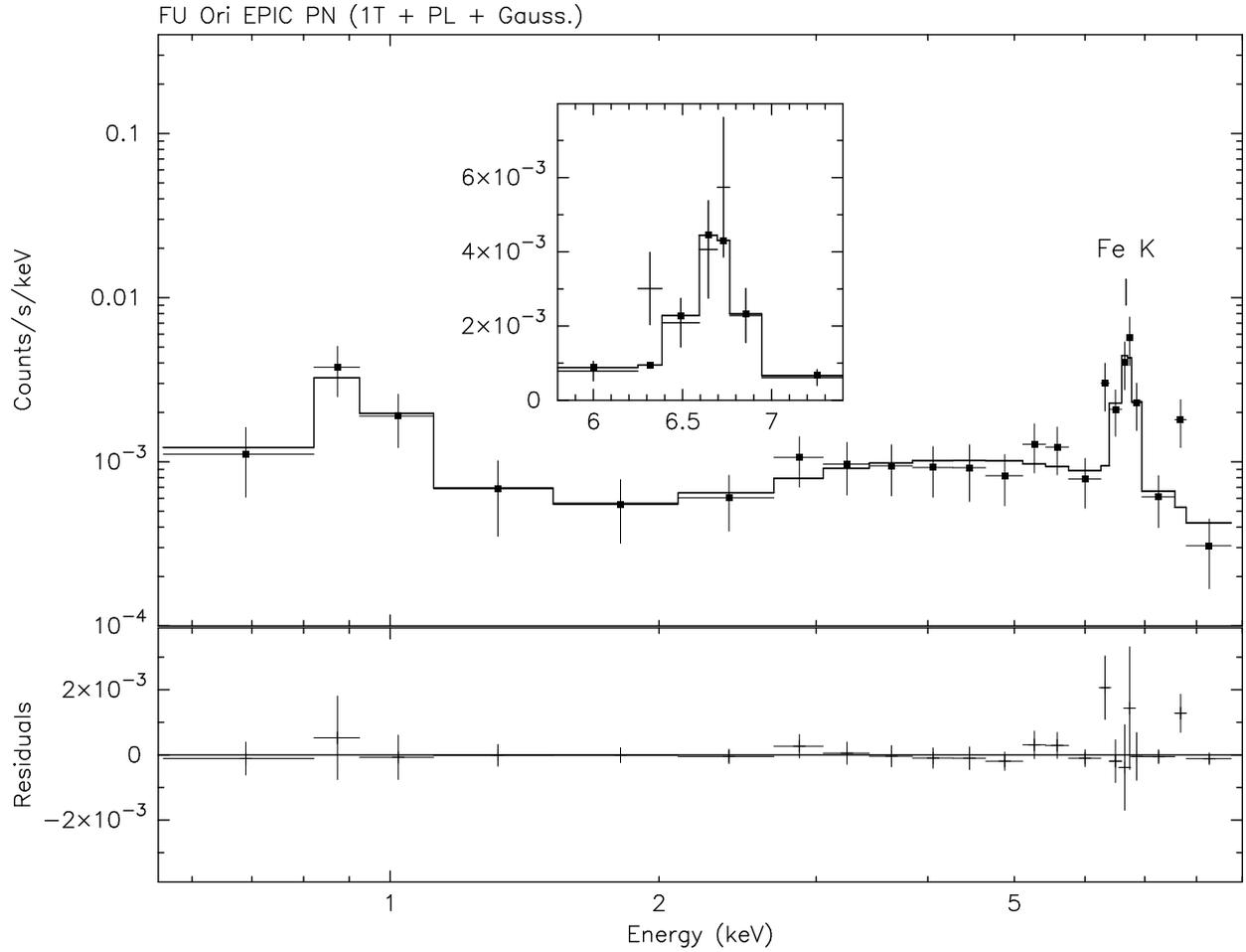}
\caption{
Same as Fig. 2 except the overlaid model
is 1T $+$ PL $+$ Gaussian Fe K line (Model D in Table 1). The Gaussian
line width was allowed to vary to achieve a best-fit.
The inset shows the fit in the vicinity of the Fe K line
on a linear axis scale. The fit attempts to account for some of
the excess below 6.5 keV and thus slightly overestimates the
true Fe K line width.
}
\end{figure}

\end{document}